\documentclass[aps,prc,showpacs,preprintnumbers]{revtex4}%
\usepackage{amssymb}
\usepackage{amsmath}
\usepackage{amsfonts}
\usepackage{graphicx}
\usepackage{bm}%
\providecommand{\U}[1]{\protect\rule{.1in}{.1in}}
\providecommand{\U}[1]{\protect\rule{.1in}{.1in}}
\begin{document}
\title{Production of exotic atoms at the CERN Large Hadron Collider (LHC)}
\author{C.A. Bertulani and M. Ellermann}
\address{Department of Physics and Astronomy, Texas A\&M University-Commerce, Commerce, TX
75429}

\begin{abstract}
We study in details the space-time dependence of the production of muonic, pionic, and other exotic atoms by the coherent photon exchange between nuclei at the Large Hadron Collider (LHC) at CERN. We show that a multipole expansion of the  electromagnetic interaction yields an useful insight of the bound-free production mechanism which has not been explored in the literature. Predictions for the spatial, temporal, and angular distribution, as well as the total cross sections,  for the production of exotic atoms are also included.
\end{abstract}
\date{\today }

 \keywords{}\maketitle

\section{Introduction}
It is undeniable that ultra-peripheral collisions  (i.e., collisions dominated by the electromagnetic interaction)  between relativistic heavy ions allows interesting physics to be probed. As far back as 1989 it was noticed that even the Higgs boson could be produced at comparable rates as in central collisions between relativistic heavy ions \cite{Pap89,Gra89}, the advantage being that ultra-peripheral collisions are cleaner in the sense that nothing else than the Higgs boson would be observed. By now, ultra-peripheral collisions are known as and excellent tool for several interesting phenomena and  have been discussed in numerous publications, reviews, and popular articles (see, e.g. \cite{BB88,BB94}).

A process of interest in ultra-peripheral collisions with relativistic heavy ions is the production of pairs in which one of the particles is captured in an orbit around one of the ions in the collider (``bound-free" pair production).  In particular the process of pair-production with capture was used as a tool to generate the first artificial anti-hydrogen atom in the laboratory \cite{Oer96}. The pioneer CERN experiment was followed by another experiment at Fermilab \cite{Bla98}. The number of anti-atoms produced was shown to be in good agreement with detailed theoretical predictions \cite{BB98}.
 
In this article we extend these studies by considering the production of exotic atoms in pp and heavy ion collisions such as those being carried out at the Large Hadron Collider (LHC) at CERN.  In our notation, an exotic atom is an  atom in which one electron has been replaced by other particles of the same charge. In contrast to previous approaches, we show that a multipole expansion of the electromagnetic field yields several insights in the production mechanism which have not been explored in the literature. Our calculations apply to the production of muonic atoms, pionic atoms, protonium, etc.  The Bohr radius for a muonic atom is much closer to the nucleus than in an ordinary atom, and corrections due to quantum electrodynamics are important. The energy levels and transition rates from excited states to the ground state of muonic atoms also provide experimental tests of quantum electrodynamics.   Hadronic atoms, such as pionic hydrogen and kaonic hydrogen, also provide interesting experimental probes of the theory of  quantum chromodynamics.  A protonium  is antiprotonic hydrogen, a composite of a negatively charged antiproton paired with a positively charged proton or nucleus.  Protonium has been studied theoretically mainly by using non-relativistic quantum mechanics, which yields predictions for its binding energy and lifetime. The lifetimes are predicted in the range of 0.1-10 microseconds. While protonium production has a very small cross section in pp collisions at CERN, the cross section is appreciable for heavy ions, e.g. for Pb-Pb collisions.  In this case, the anti-proton will be captured in an orbit around one of the Pb nuclei.  Such a system would be of large interest for understanding inter-nucleon forces in charge-conjugate channels. 

Most of our calculations will be for the production of muonic atoms and pp collisions. There is no qualitative difference for the production of other exotic atoms via the same mechanism, except for the obvious reduction of the production yields due to mass differences. The production cross sections for Pb-Pb collisions will be enhanced by a huge amount, approximately equal to $10^{10}-10^{12}$, but the details of the production mechanism are practically the same as in pp collisions. In Section II we show how a multipole expansion allows for a clear space-time description of the bound-free production mechanism. Useful approximate formulas are derived. In Section III we present our numerical results and discuss the physics properties of our results. Our conclusions are presented in Section iV.

\section{Production of exotic atoms in ion-ion collisions}
In the frame of reference of one nucleus, the time-dependent electromagnetic field generated by the other  nucleus is given by the Lienard-Wiechert potential
\begin{equation}
A_\mu({\bf r},t)=v_\mu \phi, \ \ \ {\rm with} \ v_\mu=(1,{\bf v}), \ \ \ {\rm and} \ \ \phi({\bf r},t)= {\Gamma Ze^2\over \left| {\bf R} - {\bf R}'(t)\right|},
\label{lw}
\end{equation}
where \[
{\bf r}=(x,y,z), \ \ \ {\bf R}=(x,y,\Gamma z), \ \ \ {\bf R}'=(b_x ,b_y ,\Gamma vt),\] 
and ${\bf v}$ is the relative velocity between the nuclei which we will take to lie along the z-axis.
$\Gamma$ is the relativistic Lorentz factor $\Gamma = 1/(1-v^2)^{1/2}$ (otherwise explicitly stated, here we use the units $\hbar=c=1$).

In first-order perturbation theory, the pair-production amplitude at time $t$ for a collision with impact parameter $b=\sqrt{b_x^2
+b_y^2}$ is given in terms of the transition density $\rho({\bf r})$  and the current density ${\bf j}({\bf r})$,  as
\begin{equation}
a_{1st}({\bf p}, b,t)={1\over i} \int_{-\infty}^t dt' e^{i\omega t'} \int d^3 r  j_\mu({\bf r})A^\mu({\bf r},t'), \label{a1st}
\end{equation}
where $\omega=\epsilon+m-I$, with $\epsilon$ equal to the energy of the free positive particle and $I$ being the ionization
energy of the negative captured particle. $m$ is the rest mass of either particle.
The transition current is given in
terms of the Dirac matrices $\gamma_\mu$, and the particle and anti-particle wave functions, i.e.,
 $j_\mu({\bf r}) = e{\overline \Psi_{-}} \gamma_\mu \Psi_+$,  where $\Psi_{-}$ is the
 wavefunction for the  captured negative particle and $\Psi_{+}$ that of the free positive particle.
 
The multipole expansion of $j_\mu A^\mu$ has been extensively discussed in details in Refs.~\cite{BCG03,EB02,OB09}. Replacing the Schr\"odinger by the Dirac currents in their results,  one finds that $j_\mu({\bf r})A^\mu({\bf r},t)=\sum_{\pi l \kappa} V_{\pi l\kappa}({\bf r},b,t)$, where $\pi = E,M$ and $l=0,1,2,\cdots$, and $\kappa=-l,\cdots,l$ denote the multipolarities. For electric E1 (electric dipole) and E2 (electric quadrupole) multipolarities,
\begin{equation}
V_{{\rm E1}\kappa}({\bf r},,b,t)
=   {\overline \Psi}_{-}  \left( {\bf r}\right)  (1-\gamma_0\gamma_z)  r Y_{1\kappa}\left( \hat {\bf r}\right)  \Psi_+ \left( {\bf r}\right) \frac{\Gamma Ze^2 \sqrt{2\pi/3}}{\left(
b^{2}+\Gamma ^{2}v^{2}t^2\right)^{3/2}}\left\{
\begin{array}
[c]{c}%
\mp b,\ \ (\mathrm{if}\ \ \ \kappa=\pm1)\\
\sqrt{2}vt\ \ (\mathrm{if}\ \ \ \kappa=0),\
\end{array}
\right.  \label{relE1}%
\end{equation}
\begin{equation}
V_{{\rm E2}\kappa}({\bf r},b,t)
=    {\overline \Psi}_{-} \left( {\bf r}\right) (1-\gamma_0\gamma_z)  r^{2}Y_{2\kappa}\left(
\hat{\bf r}\right) \Psi_+ \left( {\bf r}\right)   \frac{\Gamma
Ze^2\sqrt{3\pi/10}}{\left(  b^{2}+\Gamma
^{2}v^{2}t^2\right)  ^{5/2}}\left\{
\begin{array}
[c]{c}%
b^{2},\ \ \ \ (\mathrm{if}\ \ \ \kappa=\pm2)\\
\mp(\Gamma^2+1)bvt,\ \ \ \ (\mathrm{if}\ \ \ \kappa=\pm1)\\
\sqrt{2/3}\left(  2\Gamma^{2}v^{2}t^2-b^{2}\right)  \ \ \ \ (\mathrm{if}%
\ \ \ \kappa=0).\
\end{array}
\right.  \label{relE2}%
\end{equation}

In addition, for magnetic dipole (M1) \cite{OB09},
\begin{equation}
V_{{\rm M1}\kappa}({\bf r},b,t)
= \left({p_z\over m}\right) {\overline \Psi}_{-} \left( {\bf r}\right)  (1-\gamma_0\gamma_z)\gamma_z r Y_{1\kappa}\left( \hat {\bf r}\right) \Psi_+ \left( {\bf r}\right) 
 \frac{\Gamma  Ze^2\sqrt{2\pi/3}}{\left(
b^{2}+\Gamma ^{2}v^2t^2\right)^{3/2}}\left\{
\begin{array}
[c]{c}%
\pm b,\ \ (\mathrm{if}\ \ \ \kappa=\pm1)\\
0\ \ (\mathrm{if}\ \ \ \kappa=0).\
\end{array}
\right.  \label{relM1}%
\end{equation}

For the positive particle wavefunction we use a plane wave and a correction term to account
for the distortion due to the nuclear charge \cite{BB98}. The wavefunction is given by  $\Psi_{+}={\cal N}\left[{\rm v}({\bf p})\exp(i{\bf p}\cdot{\bf r})+\Psi'_+\right]$, 
where ${\cal N}(\epsilon)=\exp(\pi a_+/2)\Gamma(1+ia_+)$, with $a_+=Ze^2m/p$, with ${\bf p}$ being the proton momentum,  ${\rm v}({\bf p})$ the particle spinor and
 $\left|{\cal N}(\epsilon)\right|^2=2\pi a_+/[\exp(2\pi a_+)+1]$. 
The correction term $\Psi'_+$ is given by eq. (B5) of  ref. \cite{BB98}, which for our purposes can be written as
\begin{equation}
\Psi'_+({\bf r})={Ze^2\over 2\pi^2}{\rm v}({\bf p}) \int d^3 q e^{-i{\bf q}\cdot{\bf r}} {2\gamma^0 \epsilon +i\boldsymbol{\gamma}\cdot ({\bf q} -{\bf p}) \over ({\bf p}-{\bf q})^2(q^2-p^2)}, \label{corr1}
\end{equation}
with $\epsilon=\sqrt{p^2+m^2}$. Parts of this integral can be done analytically.

For the negative particle we use the distorted hydrogenic wavefunction  \cite{BB98} for capture to the ground state,
\begin{equation}
\Psi_-({\bf r})={1\over \sqrt{\pi}}\left({Z\over a}\right)^{3/2} \left[1-{i\over 2}\gamma^0 \boldsymbol{\gamma}\cdot \boldsymbol{\nabla}\right]{\rm u}(\epsilon_0) \exp\left({-Zr/a}\right),\label{corr2}
\end{equation}
where $u$ is the negative particle spinor, $\epsilon_0$ is the energy of the exotic atom bound-state, and $a= 1/me^2$  being the hydrogen Bohr radius. As noted in \cite{BB98},  the reason
why we need to keep the corrections in the wave functions
to first order in $Ze^2$ is because for $\epsilon \gg m$ these corrections yield a
term of the same order of magnitude as the plane-wave to the total cross section. The reason for this  
are the small distances which enter in the calculation of the
integral in Eq. \eqref{a1st}. The wavefunction corrections are essential to
account for the short distance effects properly. Such effects are not as important in the case of bound-free production of heavy particle-antiparticle pairs, but we keep them for a more accurate description of the process.

To gain insight into the pair production probability with capture and its distribution in space-time, we will compare our numerical results with  simplified calculations which neglect the above mentioned short-distance corrections of the wavefunction of the captured negative particle as well as  of the free positive particle. 
This is a reasonable approximation for small values of the free particle energy, i.e. for $\epsilon \sim m$, as we show later. This approximation  allows us to obtain the coordinate integral in $a_{1st}$ analytically, by using
\begin{equation}
F_{\lambda\kappa}({\bf p})={1\over \sqrt{\pi}}\left({Z\over a}\right)^{3/2} \int d^3 r e^{-i{\bf p}\cdot{\bf r}} r^\lambda Y_{\lambda \kappa}(\hat{\bf r}) e^{-Zr/a}=
32\sqrt{\pi} \left( {a\over Z}\right)^{5/2}  
\left\{\begin{array}[c]{c c}
\displaystyle{x_p\over (1+x_p^2)^3}Y_{1\kappa}(\hat{\bf p})   \ \ \ {\rm for} \ \ \lambda=1, \\ \; \\
       - \displaystyle{ 6(a/Z)x_p^2 \over (1+x_p^2)^4}Y_{2\kappa}(\hat{\bf p})   \ \ \ {\rm for} \ \ \lambda=2.
\end{array}\right.
\end{equation}
where $x_p=ap/Z$ and $\hat{\bf p}$ is the unit vector for the positive particle momentum direction with respect to a z-axis pointing along the beam direction.
To calculate total cross sections we keep the wavefunction corrections, Eqs. \eqref{corr1} and \eqref{corr2}. In this case, some of the coordinate integrals need to be perfomed numerically.

The exotic atom production probability density, at time $t$ and for a collision with impact parameter $b$, is obtained by  the square of expression \eqref{a1st}, i.e.,
\begin{equation}
{\cal P}({\bf p}, b,t) = \sum_{\rm spins} \left| a_{1st} ({\bf p}, b,t) \right|^2 , \label{probden}
\end{equation}
where the sum over spins is performed with standard trace techniques.  

The production probability, at time $t$ and  for a collision with impact parameter $b$, is obtained by integration over momentum,
\begin{equation}
P(b,t)=\int {d^3p\over (2\pi)^3} {\cal P}({\bf p}, b,t).
\end{equation} 
If one neglects the derivative corrections in the particle and anti-particle wavefunction, the sum over spins in eq. \eqref{probden} yields
\begin{equation}
P(b,t) = 4m\sum_{\lambda\kappa}\left| \int {d^3p \over (2\pi)^3}  \sqrt{\epsilon +m}{\cal N}(\epsilon)F_{\lambda\kappa}({\bf p})  \right|^2 \left| G_{\lambda\kappa}(b,\omega,t)\right|^2 , \label{probden3}
\end{equation}
where the small binding energy $I$ was neglected as it is small compared to the mass $m$. The functions $G_{\lambda\kappa}$ are defined as $G_{\lambda\kappa}=\int_0^t \exp\{-i\omega t'\}\{\cdots\}$, where the terms inside braces $\{\cdots\}$ are the  time-dependent terms to the right  of the positive particle wavefunctions in eqs. (\ref{relE1}-\ref{relM1}). For the M1 magnetic multipole the same equation can be used with the replacement $\sqrt{\epsilon+m} \rightarrow \sqrt{\epsilon - m}$, and multiply it by an extra factor $(p_z/m)^2$.

The cross section for bound-free pair production is obtained by integrating the production probability over all impact parameters  at $t=\infty$, i.e.,
\begin{equation}
\sigma=\int d^2 b P( b,\infty).
\end{equation}

The above integral over impact parameters diverge if the potentials of eqs. \eqref{relE1}-\eqref{relM1} are used. The reason is that these potentials are obtained from an expansion of the full Lienard-Wiechart potential of eq. \eqref{lw} which is only valid for distant collisions, i.e., when the Lorentz-modified  projectile coordinate $R'$ is larger than the Lorentz-modified internal coordinate $R$. A better approach is to use a full representation of the potential in terms of a momentum transform, as done in ref. \cite{BB98}, i.e.,
\begin{equation} 
\phi({\bf r},t)={\Gamma Ze \over 2\pi^2} \int d^3q \displaystyle{e^{i{\bf q}\cdot [{\bf R} - {\bf R}'(t)]}\over q^2} .
\end{equation}
This introduces extra integrations over the virtual momentum ${\bf q}$ increasing considerably the numerical effort \cite{BB98}.

Recently, it was shown that one can treat the effects of close collisions ($R'<R$) by using the potentials of eqs. (\ref{relE1}-\ref{relM1}) for $R'>R$ and the potentials for close collisions given by  \cite{OB09} (we will only treat the E1 case for reasons to be shown later)
\begin{equation}
V^{\mathrm{close}}_{E1\kappa}\left(  \mathbf{r},b,t\right)
= {\overline \Psi}_{-}  \left( {\bf r}\right)  (1-\gamma_0\gamma_z)  \frac{1}{r^{2}}Y_{1\kappa}^{\ast} \left(
\mathbf{\hat{r}}\right) \Psi_+ \left( {\bf r}\right) Ze^{2}\sqrt{\frac{2\pi}{3}}
 \left\{  g_{0}\left(  \Gamma\right)  +c_{\kappa}%
g_{2}\left(  \Gamma\right)  \right\}  \left\{
\begin{array}
[c]{cc}%
\sqrt{2}vt & \quad\mathrm{if\quad}\kappa=0\\
\mp b & \quad\mathrm{if\quad}\kappa=\pm1
\end{array}
\right.,
\label{VbarClose}
\end{equation}
where
$
c_{0}={2}/{5},$ $c_{\pm1}=-{1}/{5},
$
and
\begin{equation}
g_{0}\left( \Gamma\right)
=
\frac{1}{v^2}\ln\left[  \Gamma(1+v)\right], \ \ \ \ g_{2}\left(  \Gamma\right)
=
\frac{5}{4v}\left(\frac{3}{v^2}-1\right)  \ln\left[
\Gamma (1+v)\right]  -\frac{15}{2v^2}.
\end{equation}
The integrals over ${\bf r}$ now have to be carried out by using eq. \eqref{relE1} for $R'>R$ and eq. \eqref{VbarClose} for $R'<R$.  In particular, for $b=0$ one gets
\[
{1\over i} \int_{-\infty}^t dt' e^{i\omega t'} V^{\mathrm{close}}_{E1\kappa}\left(  \mathbf{r},b,t'\right)
= \delta_{\kappa,0}{\overline \Psi}_{n}  \left( {\bf r}\right)  (1-\gamma_0\gamma_z) \frac{1}{r^{2}}Y_{10}^{\ast} \left(
\mathbf{\hat{r}}\right) \Psi_p \left( {\bf r}\right) 4\sqrt{\frac{\pi}{3}}Ze^{2} \Gamma v
 \left\{  g_{0}\left(  \Gamma\right)  +c_{0}%
g_{2}\left(  \Gamma\right)  \right\} \]
\begin{equation}
\times  {1\over \omega^2} \left\{
\begin{array}
[c]{cc}
\displaystyle{\sin\left({R \omega\over \Gamma v}\right)-{R \omega\over \Gamma v} \cos\left({R \omega\over \Gamma v}\right)} & \quad\mathrm{if\quad}R\le \Gamma vt\\ \ \ \\
\displaystyle{e^{-i R \omega/ \Gamma v} \left( i -{R \omega\over \Gamma v}\right)+i e^{it \omega} (i t \omega-1)} & \quad\mathrm{if\quad}R>\Gamma vt
\end{array}
\right.,
\label{VbarClose2}
\end{equation}
For $\Gamma \gg 1$, $g_0(\Gamma)+c_0g_2(\Gamma) =2\ln(2\Gamma)-15/2$. When $t=\infty$ only the upper term in the last part of this equation contributes ($R<\infty$) to the production probability.

We use the above separation of close and distant collisions  to make an estimate of their relative contributions to the total cross section. To obtain the total cross sections we use the formalism developed in ref. \cite{BB98} which allows a better reduction of the integrations in momentum space. Adapting their results for ion-ion collisions, we get

\begin{equation}
\sigma =\frac{256\pi Z^{7}e^{4}}{3 a^{5}}\int_{m-I}^{\infty }d\epsilon \ \left| {\cal N}(\epsilon)\right|^2 {p\over \omega}\int_{0}^{\infty } dq\;\frac{q(q^2+\omega^2/\Gamma^2 v^2)}{(q^2-\omega^2/\Gamma^2v^2)^{2}}\frac{\left( p^{2}+3%
q^2+3\omega^2/v^2\right) \left( 3 p^{2}+q^{2}+\omega^2/v^2\right) }{\left( q^2+\omega^2/v^2
- p^{2}\right) ^{6}}. \label{4.3}
\end{equation}

Notice that we keep the relative velocity $v$ between the
colliding nuclei in all formulas, although $v\sim c$. This is necessary
because sometimes important combinations of $1$ and $v$ conspire and combine into Lorentz factors 
$\Gamma=(1-v^2)^{-1/2}$ in subsequent steps of the calculations. 
 
\section{Results}

\begin{figure}[t]
\centerline{
\includegraphics[width=70mm,keepaspectratio]{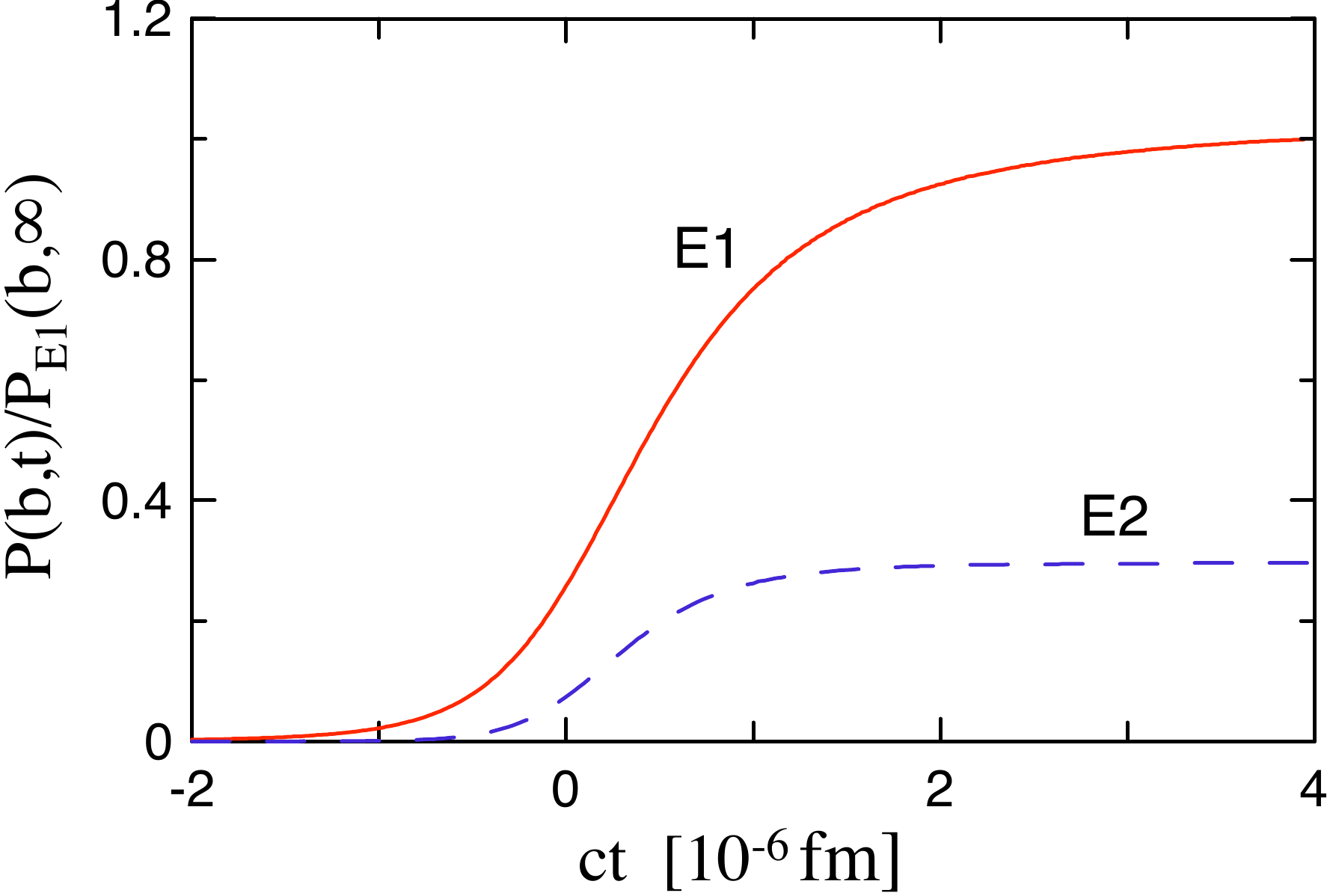}
} \caption{\label{fig1}    Relative probability $P_{E\lambda}(b,t)/P_{E1}(b,t=\infty)$,  for  production of a muonic atom in pp collisions at the LHC with an impact parameter $b=a_{muon}=255$ fm, where $a_{muon}$ is the Bohr radius for a muonic atom.}\label{tdep}
\end{figure}

 At the Large
Hadron Collider (LHC) at CERN the Lorentz gamma
factor  in the laboratory frame, $\Gamma_{lab}$, is 7000 for
p-p, 3000 for Pb-Pb collisions.  The relationship between the Lorentz contraction factor associated with the relative
velocity between the colliding nuclei, and the collider energy per nucleon, $E/A$, in GeV, 
is given by $\Gamma=2(\Gamma_{lab}^2-1)\simeq2(1.0735 E/A)^2$. This means that for the production of exotic atoms we have effectively
$\Gamma \simeq 10^8$ for p-p and $\Gamma \simeq 10^7$ for Pb-Pb collisions.

In figure \ref{tdep} we plot the relative probability $P_{E\lambda}(b,t)/P_{E1}(b,t=\infty)$,  for  production of a muonic atom in pp collisions at the LHC with an impact parameter $b=a_{muon}=255$ fm, where $a_{muon}$ is the Bohr radius for a muonic atom. The time scale is in units of $10^{-8}$ fm/$c$. The probability is normalized so that it is equal to the asymptotic value of the E1 multipolarity. In absolute values the probabilities are very small, justifying the use of perturbation theory.  The E2 probability tends to its asymptotic value faster than  E1. This can be understood from Eqs. \eqref{relE1} and \eqref{relE2} as the time-dependence of the E2-field is determined by a higher inverse power of time. The asymptotic value of the E2 probability is about a factor 4 smaller than the E1 case. As a function of the impact parameter, the E2 probability decreases with an additional $1/b^2$ dependence,  as compared to E1. This yields cross sections for pair-production with capture due to the E2 field being much smaller than that with E1.
The probabilities and cross sections are also much smaller in the case of the M1 multipolarity, due to the factor ($p_z/m$) in Eq. \ref{relM1}. This is also substantiated by  the approximation in Eq. \eqref{probden3}, which has a factor $\sqrt{\epsilon-m}$ instead of  the  $\sqrt{\epsilon+m}$ which appears in the electric multipole cases. For low positive particle momentum this leads to a further suppression of the M1 multipolarity. 

We thus conclude that the E1 multipolarity alone is responsible for most part of the bound-free pair-production probability. This comes as no surprise because of the very large $\Gamma$ factor in the frame of reference of the exotic atom. The spatial distribution of the time-dependent  field is compressed as a pancake-like object with a spatial width $\Delta z = c\Delta t = b/\Gamma$. For $b=200$ fm this is equal to $\Delta z \simeq 10^{-6}$ fm for the LHC. Even for very large impact parameters, e.g., $1 \AA =10^5$ fm, $\Delta z$ is small compared to the nuclear sizes. The E2 field is a measure of the ``tidal" force, proportional to the spatial spreading of the electric field \cite{BB88}. This ``tidal" effect  becomes larger at smaller velocities, when the field lines are not as compressed.   The large value of $\Gamma$ also leads to a complete dominance of the $\kappa=\pm 1$ component of the E1 field in eq. \eqref{relE1}.

The above discussion implies that in order to calculate pair-production with capture in ultra-peripheral collisions at the LHC once only needs to consider the E1 field, with $\kappa= \pm 1$, in eq. \eqref{relE1}. The asymptotic production probability amplitude, Eq. \eqref{a1st},  for a given impact parameter $b$,  is then given by
\begin{equation}
a_{\kappa=\pm 1}({\bf p}, b)= 2\kappa i {Ze^2 \over vb}\sqrt{2\pi\over 3} \xi K_1(\xi)\int d^3 r  {\overline \Psi}_{n}  \left( {\bf r}\right)  (1-\gamma_0\gamma_z) r Y_{1\kappa}\left( \hat {\bf r}\right) \Psi_p \left( {\bf r}\right)   , \label{a1st2}
\end{equation}
where $\xi=\omega b/\Gamma v$ and $K_1$ is the modified Bessel function of first order. Note that  $\xi K_1(\xi)\simeq 1$ for $\xi \lesssim 1$. For $\xi >1$, $\xi K_1(\xi)$  drops to zero exponentially. This means that the production probability drops as $1/b^2$ up to $b_{max} \approx \Gamma \hbar c/m \approx  10^{10} ({\rm MeV.fm})/mc^2$ for pp collisions at the LHC. For production of muons and pions with capture this means $b_{max} \approx 10^8$  fm, whereas for proton-antiproton with capture, this means $b_{max} \approx 10^7$ fm. 

\begin{figure}[t]
\centerline{
\includegraphics[width=85mm,keepaspectratio]{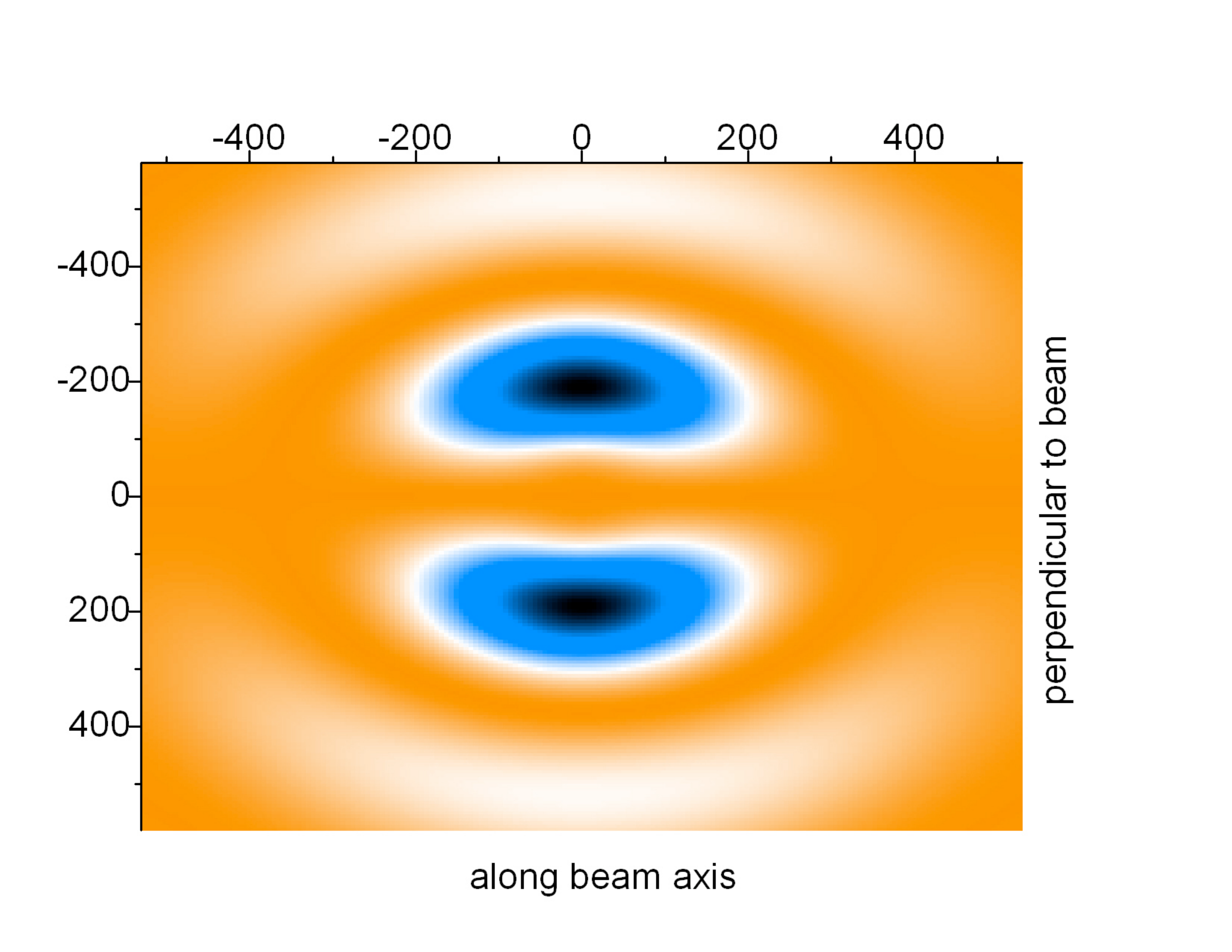}
\includegraphics[width=85mm,keepaspectratio]{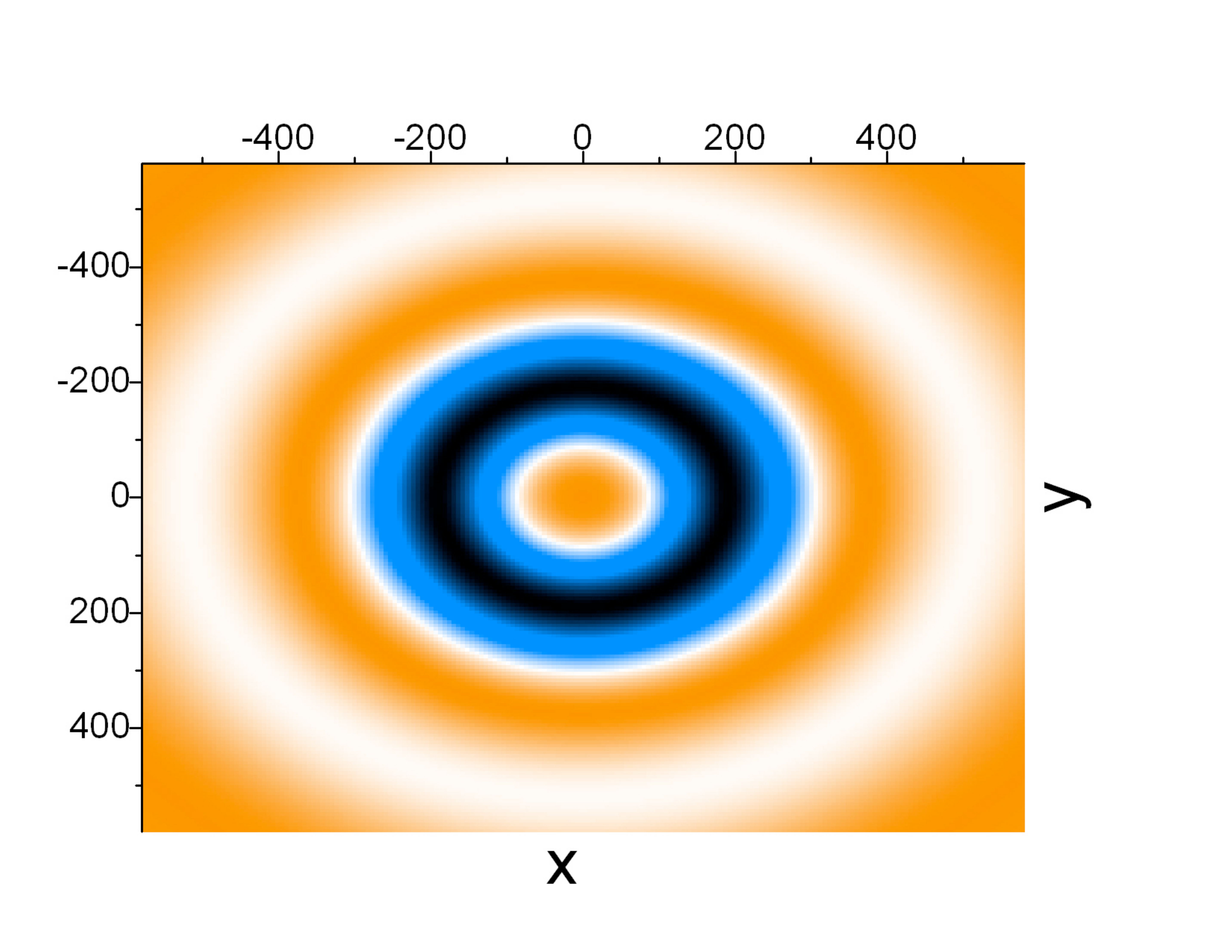}
} \caption{ Left: Coordinate distribution in the plane along the beam axis of the  probability for production of muonic atoms with capture in the K-shell with pp collisions at the LHC.  The impact parameter is chosen as $b=a_{muon}=255$ fm and is along the vertical axis. Right: Same as the previous figure, but in the plane perpendicular to the beam axis. The impact parameter vector lies along the x direction.}\label{figang1}
\end{figure}

In the left panel of Fig.  \ref{figang1}  we show the coordinate space distribution for production of muons with capture in the K-shell in pp collisions at the LHC.  The distribution is shown in a plane containing the beam direction and the impact parameter vector.  The calculation is done for the E1 multipolarity and for an impact parameter $b=a_{muon}$. The positive muon energy is taken as $\epsilon=1.1 m_\mu$ and its direction of emission is chosen as $100^\circ$ when measured along the direction of motion of the muonic atom. One notices that the production mechanism is more efficient in regions perpendicular to the beam axis. The darker areas are representative of larger production rates in coordinate space. The right figure shows the same result, but as seen in the plane perpendicular to the beam axis. In this case, the impact parameter vector lies along the x-direction. We observe that the probability density is largest within a torus-like region with a radius $r\approx a_{muon}$ from the origin of the muonic atom, with the torus axis along the beam direction. 

We now use Eq. \eqref{VbarClose2} to obtain the production probability of muonic atoms at the LHC at zero impact parameter. We find that the probability is 5 orders of magnitude smaller than that with $b=a_{muon}$. We easily understand this result by inspection of the equation \eqref{VbarClose2}.
When $t=\infty$ only the upper term in the last part contributes ($R<\infty$) to the production probability. This term oscillates harmonically as a function of $\omega R/\Gamma v\approx 2mR/\Gamma v$. For $R$ along the z-direction  this variable varies as $2mc^2 z/\hbar c$. As the largest contribution to the integral arises for $z\approx a_{muon}$, this variable is much larger than the unity, causing the harmonic functions to oscillate wildly, and leading to a small value of the integral over coordinates  (matrix element).

\begin{figure}[t]
\centerline{
\includegraphics[width=70mm,keepaspectratio]{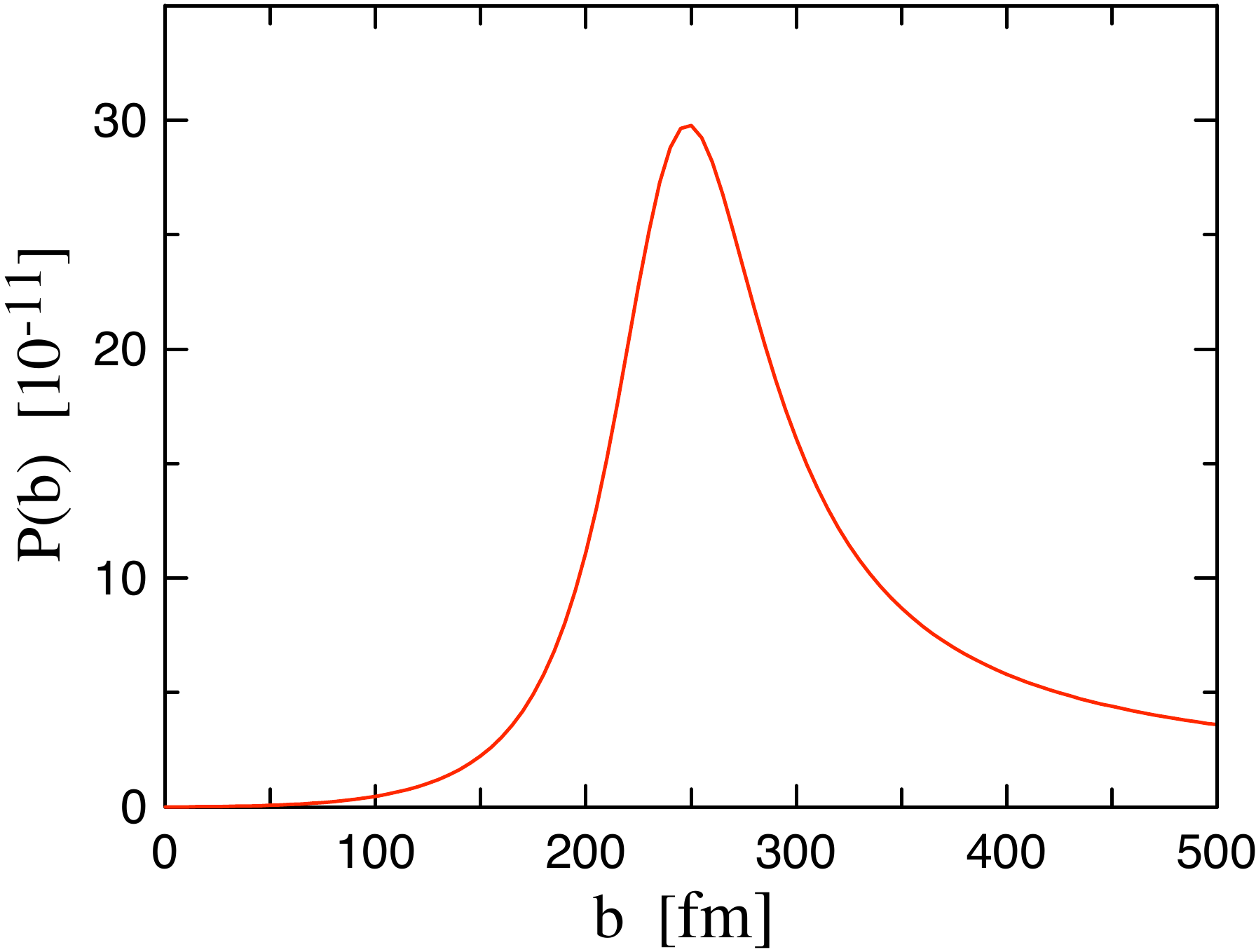}
} \caption{ Probability of muonic atom production in pp collisions at the LHC  as a function of the impact parameter. }\label{bdep}
\end{figure}

We now perform an estimate of the contribution of collisions with small impact parameters by using Eq. \eqref{VbarClose2} and integrating it numerically over time, space, and impact parameters, for several energies and angles of the emitted positive muon. To obtain these estimates, we have used the functions $\Psi_-$ and $\Psi_+$ without the derivative corrections. The integrals are performed for $b<a_{muon}$. We compare with the results obtained using  Eq.   \eqref{a1st2} for $b\ge a_{muon}$. We find that the impact parameter region $b>a_{muon}$ contributes at least a factor 100 more to the cross section than the region with $b<a_{muon}$.  Such findings are confirmed by an integration over the free positive muon energy. This is also confirmed using Eqs. \eqref{a1st2} and \eqref{VbarClose2} and  $\Psi_-$ and $\Psi_+$ without derivative corrections. Our numerical results are shown in figure \ref{bdep}. We clearly see that the small impact parameters, $b\lesssim a_{muon}$ yield a much smaller probability than $b\gtrsim a_{muon}$.

\begin{figure}[t]
\centerline{
\includegraphics[width=70mm,keepaspectratio]{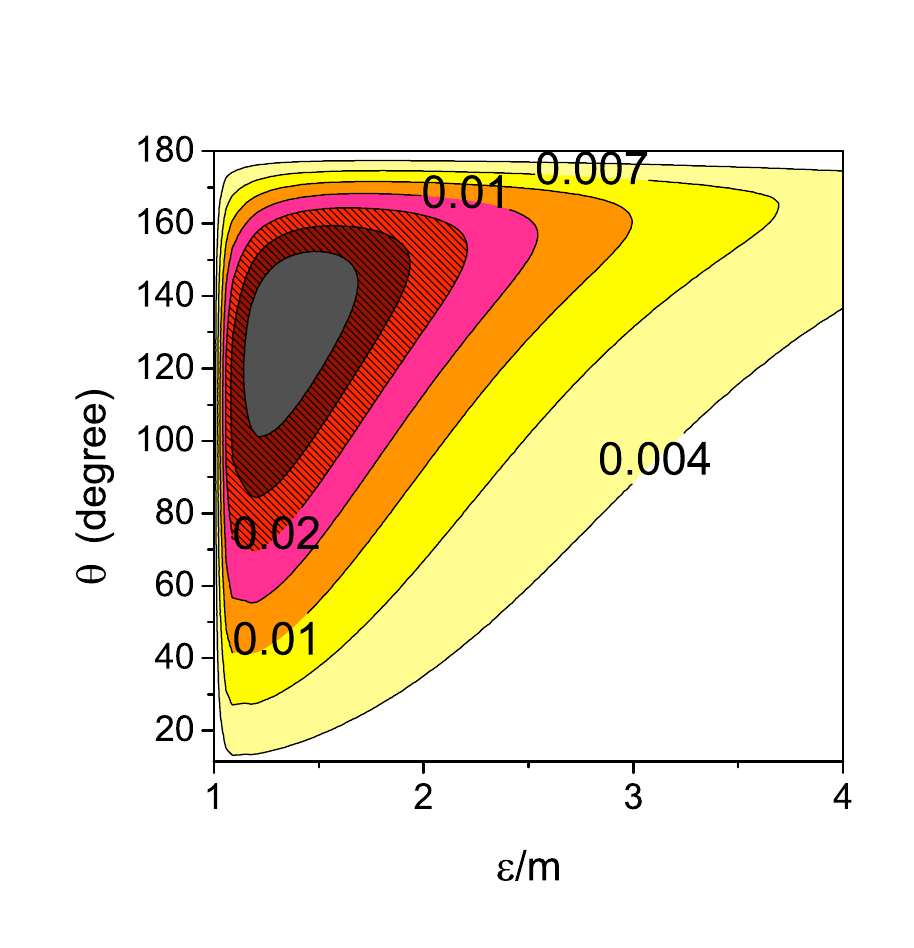}
} \caption{\label{fig1} Contour plot with the angular distribution of the positive muon when the negative muon is captured by a proton at the LHC, as a function of the angle that the free muon has with the direction of motion of the muonic atom and of the energy of the free positive muon. }\label{figang}
\end{figure}

In Fig. \ref{figang}, we plot the angular distribution of positive muon when the negative muon is captured by a proton at the LHC as a function of the angle that the free muon has with the direction of motion of the muonic atom and of the energy of the free positive muon. The impact parameter is chosen as $b=a_{muon}$. The units in the contour plot are arbitrary, with the darker areas being the region of highest probability. The plot shows that  the higher the electron energy is, the more backward peaked the distribution becomes. 
In the frame of reference of the atom, the angular distribution of the positive particles is
backward peaked, along the beam axis, opposite in the direction of motion of the nucleus capturing  the muon. The higher the positive muon energy is, the
more backward peaked the distribution becomes. The most probable energy of the free muon is non-relativistic, i.e., $\epsilon \simeq  m$.  For such low energy muons the opening angle for emission of the free muon is larger.
 In a collider, a Lorentz transformation of
these results to the laboratory frame implies that all particles are seen along the beam
direction, within an opening angle of order of $1/\Gamma$.  That is, all positive particles are seen along the same direction as the beam.

Using Eq. \eqref{4.3} we calculate the cross sections for production of exotic atoms in pp and Pb-Pb collisions at the LHC. Our results are shown in Table I. Whereas the cross sections are small for pp collisions, they are  by no means negligible for Pb-Pb collisions. This is due to the factor $Z^7$ in Eq.  \eqref{4.3}, although a reduction of this $Z$ dependence arises from the distortion factor ${\cal N}$.

\begin{table}
\begin{center}%
\begin{tabular}
[c]{llllll}\hline\hline
exotic atom  & \ \ \ pp & \ \ \ Pb-Pb
\\\hline
hydrogenic & \ \ \ 63.4 pb & \ \ \ 132 b\\
muonic &  \ \ \ $44.8 \times 10^{-4}$ pb & \ \ \ 0.16 mb\\
pionic &  \ \ \ $21.3 \times 10^{-4}$ pb & \ \ \ 0.09 mb\\
kaonic & \ \ \ $1.3 \times 10^{-4}$ pb & \ \ \ 4.3 $\mu$b\\
$\rho$-atom & \ \ \ $0.51 \times 10^{-4}$ pb& \ \ \ 1.3 $\mu$b\\
protonium & \ \ \ $0.09 \times 10^{-4}$ pb & \ \ \ 0.3 $\mu$b\\\hline\hline
\label{t2} &  &  
\end{tabular}

\caption{Cross sections for production of exotic atoms in pp and Pb-Pb collisions at the CERN Large Hadron Collider (LHC).}
\end{center}
\end{table}

\section{Conclusions}

 Because of the  very strong
electromagnetic fields of short duration, new 
and interesting physics arise at the LHC.
We have studied the space-time dependence of the cross sections for
 production of exotic atoms in ultraperipheral collisions 
at the LHC.  We have considered the case of muonic, pionic and anti-protonic (or protonium) atoms.
A very transparent and simple formulation is obtained using
the lowest-order corrections for the positron and electron wavefunctions and a multipole expansion of the electromagnetic field separated in the regions of small (i.e., $b$ smaller than the Bohr radius) and of large impact parameters.
Whereas most of our discussion and conclusions have been  based on calculations of muonic atoms, the qualitative aspects will not change for pionic, protonic, or other exotic atoms, except for the magnitude of the cross sections. The space-time and impact parameter dependences will also be similar for Pb-Pb collisions.

We considered the  capture of negative particles in the K-shell of either the proton of the Pb nucleus. This is the largest contribution to the capture cross section. Inclusion of capture to all other shells will  increase by about  20\% of the our calculated cross sections, according to early estimates for electron-pair production with capture \cite{BB88}. Depending on the relevance of this process to peripheral
collisions with relativistic heavy ions and on the accuracy attained by the experiments,
further theoretical studies on the capture to higher orbits could be necessary. The distortion of the wavefunctions due to the nuclear size should also be considered.
 Due to their small charges, pp collisions will not yield a measurable amount of exotic atoms, but in Pb-Pb collisions we expect an  abundant number of events of production of exotic atoms. This will open opportunities to study the properties of exotic atoms and their decay widths.
 
\medskip 
\acknowledgments{This work was partially supported by the U.S. DOE grants DE-FG02-08ER41533
and DE-FC02-07ER41457 (UNEDF, SciDAC-2), and the Research Corporation.}


\begin{thebibliography}{99}  

\bibitem{Pap89}
E. Papageorgiu, Phys. Rev. D 40, 92 (1989).

\bibitem{Gra89} 
M. Grabiak, B.M\"uller,W. Greiner, G. Soff, and P. Koch, J. Phys.
G 15, L25 (1989).                                                                                             

\bibitem{BB88}
C.A.Bertulani and G.Baur, Phys. Rep. 163, 299 (1988).

\bibitem{BB94} C.A. Bertulani and G. Baur, Physics Today, March 1994, p. 22.

\bibitem{Oer96} G. Baur et al., Phys. Lett. B 368, 251 (1996).

\bibitem{Bla98} G. Blanford, Phys. Rev. Lett. 80, 3037 (1998).

\bibitem{BB98}  C.A. Bertulani and G. Baur, Phys. Rev. D58, 034005 (1998).

\bibitem{BCG03} C.~A.~Bertulani, C.~M.Campbell, and T. Glasmacher,
Comp. Phys. Comm. 152, 317 (2003).

\bibitem{EB02}
H. Esbensen and C. A. Bertulani, Phys. Rev. C 65, 024605 (2002).

\bibitem{OB09} K.~Ogata and C.~A.~Bertulani, Prog. Theo. Phys. 121, 1399 (2009); Prog. Theo. Phys., in press.

\end{thebibliography}
\end{document}